\documentclass[twocolumn,aps,prba,superscriptaddress]{revtex4-1}
\usepackage[utf8]{inputenc}
\usepackage{amsmath}
\usepackage{amssymb}
\usepackage{graphicx}
\usepackage{booktabs}
\usepackage{physics}
\usepackage{siunitx}
\usepackage[format=plain]{caption}
\usepackage{subfig}
\usepackage[margin=0.9in]{geometry}
\usepackage{textcomp}
\usepackage{gensymb}
\usepackage{sidecap}
\usepackage[english]{babel}
\usepackage[autostyle]{csquotes}
\usepackage{braket}
\usepackage{array}
\usepackage{color}
\usepackage{multirow}

\begin{document}
\justifying
\captionsetup{justification=Justified,}

\title{Million-atom simulation of the set process in phase change memories at the real device scale}
\author{Omar Abou El Kheir}
\affiliation{Department of Materials Science, University of Milano-Bicocca, Via R. Cozzi 55, I-20125 Milano, Italy}
\author{Marco Bernasconi}
\affiliation{Department of Materials Science, University of Milano-Bicocca, Via R. Cozzi 55, I-20125 Milano, Italy}

\begin{abstract}
Phase change materials are exploited in several enabling technologies such as storage class memories, neuromorphic devices and memories embedded in microcontrollers. A key functional property for these applications is the fast crystal nucleation and growth in the supercool liquid phase. Over the last decade, atomistic simulations based on density functional theory (DFT) have provided crucial insights on the early stage of this process. These simulations are, however, restricted to a few hundred atoms for at most a few ns. More recently, the scope of the DFT simulations have been greatly extended by leveraging on machine learning techniques. In this paper, we show that the exploitation of a recently devised neural network potential for the prototypical phase change compound Ge$_2$Sb$_2$Te$_5$, allows simulating the crystallization process in a multimillion atom model at the length and time scales of the real memory devices. The simulations provide a vivid atomistic picture of the subtle interplay between crystal nucleation and crystal growth from the crystal/amorphous rim. Moreover, the simulations have allowed quantifying the distribution of point defects controlling electronic transport, in a very large crystallite grown at the real conditions of the set process of the device.
\end{abstract}


\maketitle  
\section{Introduction}

Chalcogenide glasses, such as the prototypical Ge$_2$Sb$_2$Te$_5$  (GST) compound, are exploited in  non-volatile electronic memories, either standalone \cite{wuttig2007phase,noe2017phase,fantini} or embedded in microcontrollers, \cite{Cappelletti_2020,Redaelli2022} and in devices for in-memory and neuromorphic computing \cite{kuzum2012nanoelectronic,tuma2016stochastic,sebastianNanotech}.

These applications rely on a fast and reversible transformation between the crystalline and amorphous phases induced by Joule heating \cite{wuttig2007phase}. The two phases feature a difference in resistivity by three orders of magnitude that allows encoding a binary information which can be read out by a measurement of resistance at low bias \cite{wuttig2007phase}. Amorphization of the crystal via melting (reset) and recrystallization of the amorphous phase (set) are achieved by applying current pulses at higher bias \cite{wuttig2007phase,noe2017phase}. Partial recrystallization during set or modulation of the size of the amorphous region during reset are also possible to obtain intermediate discrete or analogic resistance levels for in-memory and neuromorphic computing \cite{sebastianNanotech}.

The crystallization kinetics is thus a key functional property determining the programming time of the memory and as such it has been the subject of extensive experimental and theoretical studies over the years \cite{zhang2019designing}.
Experimental works have shown that GST exhibits a nucleation driven crystallization in which several overcritical crystalline nuclei form and then slowly grow \cite{salinga2013}.  Other materials in this class, such as AgInSbTe alloys \cite{salinga2013}, feature instead a growth driven crystallization which mostly proceeds at the interface between the amorphous inclusion and the crystalline matrix with no or marginal crystal nucleation. This distinction depends, however, on the temperature and on the size of the amorphous region which is ruled by the device architecture. 

On the other hand, molecular dynamics (MD) simulations based on density functional theory (DFT) provided crucial insights on the early stage of nucleation and growth in models containing a few hundreds of atoms \cite{elliot,PhysRevLett.107.145702,kalikkacrys3,ronneberger2015crystallization,ronneberger2018crystal,GST124,rao2017reducing,elliott2017}.  Most recently, machine learning techniques allowed extending the scope of DFT methods by providing interatomic potentials fitted on a DFT database that enable large scale simulations of phase change materials  \cite{SossoNN,sosso2013,sosso2015,mocanu2018modeling, dragoni2021mechanism,LEE2020109725,DeringerNatureEl,sun2023,Robertson2023,omar2024,Li2024,baratella2024}. In particular, interatomic potentials generated with neural network (NN) methods enabled the simulations of a few thousands of atoms for tens of ns that provided information on the crystal growth velocity in a wide range of temperatures for GeTe \cite{sosso2013,gabardi2017} and GST \cite{omar2024}.  Very recently, a Gaussian Approximation Potential (GAP) for GST  allowed simulating the reset process (crystal melting) in a 532,980-atom model at a length scale close to that of the real memory device \cite{DeringerNatureEl}. This is, however, the fast process in the memory programming that requires MD simulations lasting a few tens of ps.

In this work, we show that it is possible to simulate the slow process in memory operations, i.e. the crystallization of the amorphous phase (set) at the length and time scales of the real memory devices, by exploiting the efficient implementation on graphical processor units (gpus) of the DeePMD code \cite{wang2018deepmd,PhysRevLett.120.143001,lu202186} that we used previously to generate a NN potential for GST \cite{omar2024}.
MD simulations of models with 2.795 million atoms have allowed us to visualize at the atomic level the recrystallization of an amorphous dome embedded in a crystalline matrix on the time scale of a few ns.

\section{Results and Discussion}

\begin{figure*}[ht!]
 \centering
\includegraphics[width=0.8 \textwidth, keepaspectratio]{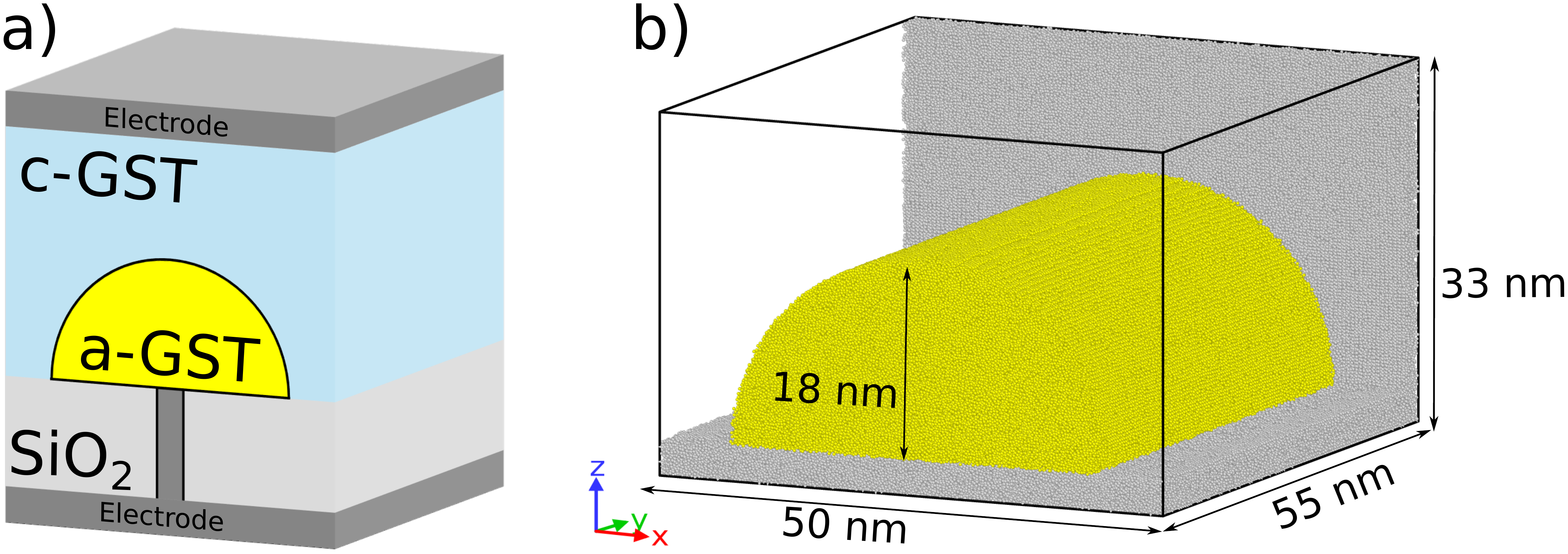}

\includegraphics[width=0.8 \textwidth, keepaspectratio]{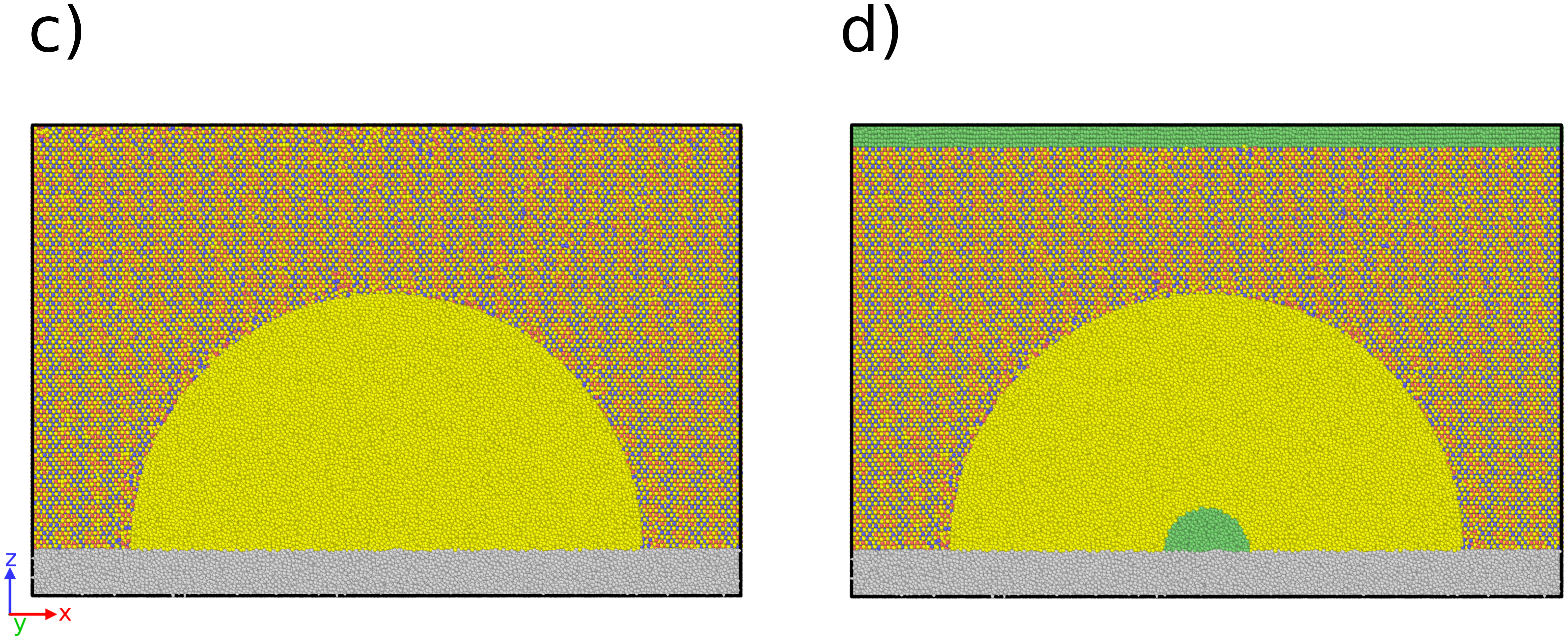}

 \caption{
 a) Sketch of the active region of a PCM in the  Wall architecture \protect\cite{Laurin2023}. A semi-cylindrical region of a-GST is embedded in a crystalline matrix (c-GST) and confined by dielectrics (SiO$_2$)
 in the $y$ direction and by  the top and bottom electrods in the $z$ direction. The heater in contact with a-GST is also shown. b)  Atomistic model of the PCM cell with 3D periodic boundary conditions. 
 The gray regions correspond to the confining materials, the yellow region is the active amorphous semi-cylindrical dome, embedded in a crystalline matrix (not shown). c) 
A snapshot of the model with atomic resolution. 
d) A snapshot of the model highlighting (in green) the atoms in the upper part of the model thermostatted at 300 K and the atoms inside the dome thermostatted at 880 K (see text). The Ovito \cite{ovito} tool was used for the visualization and the generation of all atomic snapshots of this article.}
 \label{WallPCM}
\end{figure*}

We modeled the active region of a phase change memory (PCM) device in the Wall architecture \cite{Arnaud}  sketched in Figure \ref{WallPCM}a which is used for embedded applications \cite{Laurin2023}.  In place of the dielectrics and electrodes that confine the GST film, we used a thin film  (3 nm thick) of still amorphous GST (a-GST) whose atoms were frozen at zero temperature to mimic the high thermal stability of the materials surrounding the active region. The simulation unit cell with 3D periodic boundary conditions is shown in Figure \ref{WallPCM}b. The light gray regions correspond to frozen a-GST atoms that mimic the surroundings of the active GST film which consists in turn of a semi-cylindrical amorphous dome embedded (yellow region) in a crystalline matrix (not shown). The radius of the amorphous dome and the thickness of the GST film along the $z$ and $x$ directions shown in Figure \ref{WallPCM}b have values typical of the Wall architecture for embedded memories \cite{Boniardi2014}. The model contains 2.323 million  mobile atoms and 0.472 million atoms in the frozen layers. The semi-cylindrical amorphous dome contains 786,412 atoms. A snapshot of the model with atomic resolution is shown in Figure \ref{WallPCM}c.

To generate the model of Figure \ref{WallPCM}b-c, we started from cubic GST at the theoretical density  of 0.0309 atom/\AA$^3$ \cite{omar2024}. The amorphous confining layers and the active amorphous dome were then generated by heating these regions at 1400 K for 20 ps and then quenching to 600 K in 100 ps, while all the other atoms were always thermostatted at 600 K.   The amorphous confining layers were then frozen in all the subsequent simulations. 

For the  simulation of the crystallization, we must consider that thermal conductivity $\kappa$ allows the latent heat released during the crystallization to diffuse away from the crystal growth front. In the crystalline cubic phase, $\kappa$ is mostly due to phonons (0.4 W/mK) while electrons contribute for about 10 $\%$ (0.04 W/mK) \cite{LeeWong}. In the supercooled liquid phase at temperatures above the metal-insulator transition \cite{cobelli2021,baratella2022}, the contribution of electrons to $\kappa$ is supposed to be higher, although we do not have a reliable estimate for this quantity. In our simulation, in the lack of the explicit modeling of electrons, the thermal conductivity is  sustained only by phonons. The crystallization of the amorphous dome was then simulated by controlling temperature in three different manners as described below. 

First, we performed a simulation by applying two independent thermostats for atoms in the crystalline matrix and in the amorphous region at the same target temperature of 600 K. This protocol (simulation A) allows  us to efficiently remove the latent heat released during the crystallization.  In a second simulation (simulation B), we considered the other extreme in which the latent heat is removed only by the upper electrode. This is mimicked by applying a thermostat at 600 K only to a thin crystalline film 3 nm  thick in contact with the frozen layers mimicking the upper electrode,  after having equilibrated both the embedding crystal and the embedded amorphous dome at 600 K. In a third and more realistic simulation (simulation C), we applied two thermostats at two different temperatures. One thermostat as above is applied only to a thin crystalline film  1.5 nm thick in contact with the frozen layers mimicking the upper electrode and set to 300 K. The second thermostat is applied to a small semi-cylindrical region with radius 30 {\AA } in the lower part of the dome (see Figure \ref{WallPCM}d). The second thermostat is set to 880 K to mimic the hot spot generated by the heater underneath, as it was shown by finite-elements simulations of the operation of the device in Ref. \cite{Baldo}.

\begin{figure*}[ht!]
 \centering
\includegraphics[width=0.4\textwidth, keepaspectratio]{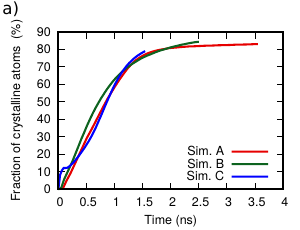}
\includegraphics[width=0.4\textwidth, keepaspectratio]{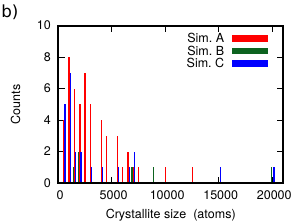}
 \caption{a) Fraction of atoms in the active region that are recrystallized as a function of time in the PCM model of Figure \ref{WallPCM}c, under different conditions (simulations A/B/C, see text). b) Distribution of the size of crystallites nucleated inside the dome.}
 \label{CrystallineAtoms}
\end{figure*}

\begin{figure*}[]
 \centering
\includegraphics[width=0.8\textwidth, keepaspectratio]{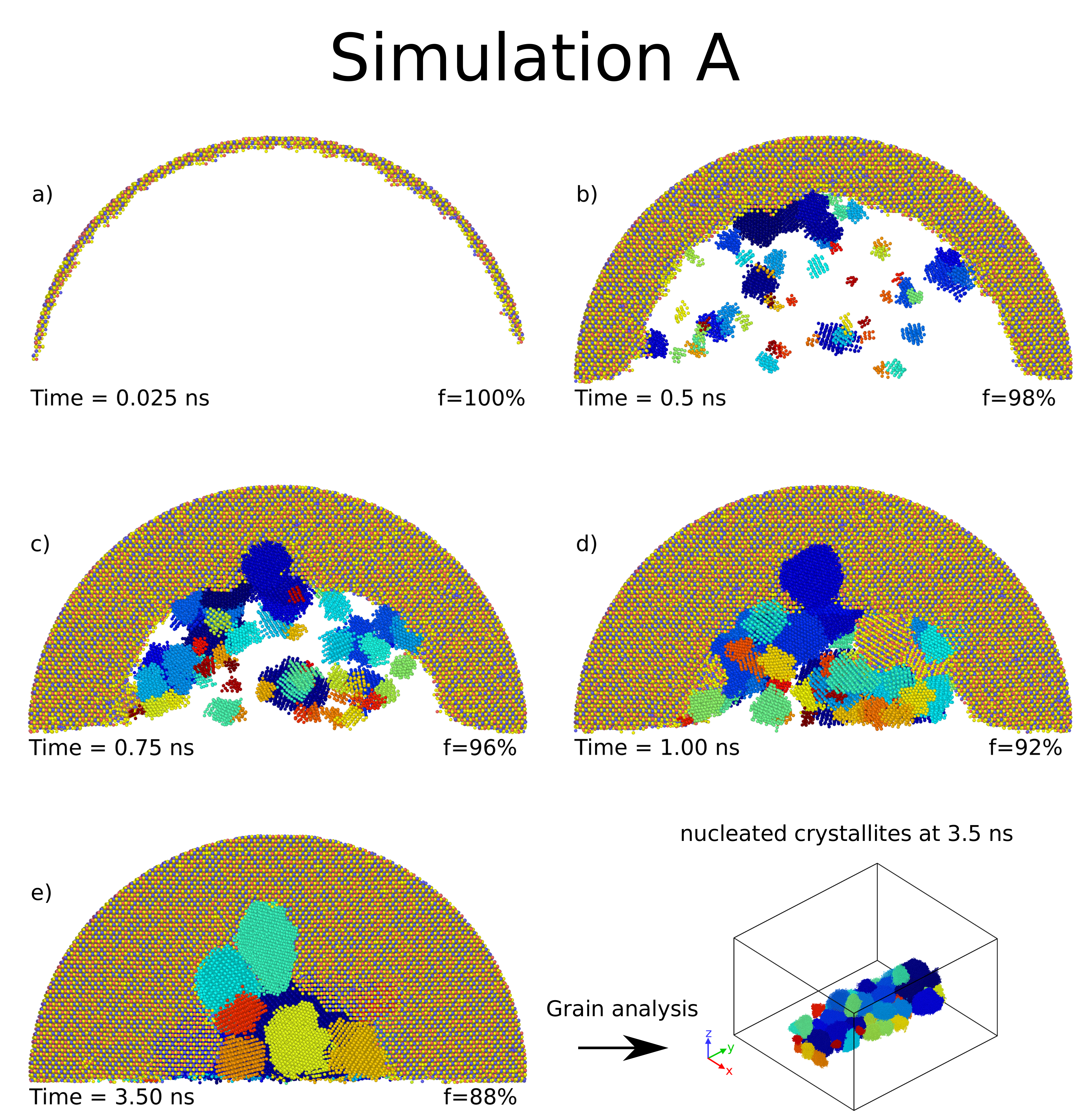}
 \caption{ a-e) Snapshots of the recrystallization of the amorphous region at different times (in ns) with the dome thermostatted at 600 K (simulation A). Only recrystallized atoms that were originally present in the amorphous semi-cylindrical dome in Figure \ref{WallPCM}b-c are shown. Atoms belonging to the largest crystallite growing form the outer amorphous-crystal interface are shown with the same colors of Figure \ref{WallPCM}c. At the end of the simulation, the largest crystallite contains 541,233 atoms. The smaller crystallites nucleated inside the dome are shown each one in a different color. The fraction of crystallized atoms (number f) belonging to the outer crystallite growing from the rim is given in each panel.  A side view of the smaller crystallites (grains) embedded in the largest one is shown on bottom right. These nucleated crystallites touch the confining walls on both sides along $y$ (see Figure \ref{WallPCM}).}
 \label{Snapshots}
\end{figure*}

We first analyze the results of simulation A with the dome thermostatted at 600 K. The evolution in time of the fraction of the recrystallized atoms is shown in Figure \ref{CrystallineAtoms}a, while snapshots of the system at different times are shown in Figure \ref{Snapshots}. To identify the crystalline nuclei, we used the local order parameter Q$_4^{dot}$  \cite{q4} that we considered in our previous work on the crystallization of bulk GST \cite{omar2024}. Most of the crystallization takes place at the crystal-amorphous rim which moves inwards with time, but crystal nucleation is also present inside the amorphous dome.

The largest crystallite grown from the amorphous-crystal interface contains 541,233 atoms which amount to 88 $\%$ of the total number of recrystallized atoms. The distribution of the size of the smaller crystallites nucleated inside the dome is shown in Figure \ref{CrystallineAtoms}b. About 75\% of the atoms crystallizes rapidly in the first 1.5 ns, then a slow down takes place due to the interaction among  several nuclei, about 55, formed within the dome.  The fraction of crystallized atoms then increases very slowly and reaches a plateau around 80\% at 3.5 ns. Most of the crystallites nucleated inside the dome contains less than 3000 atoms, but a few grows up to  12.000 atoms  (see Figure \ref{CrystallineAtoms}b). These crystallites are expected to undergo a coarsening on a longer time scale as was shown in Ref. \cite{omar2024}. 

The crystal growth velocity $v_{\rm g}$ of the crystallite growing from the interface has been computed from the evolution of the radius of semi-cylindrical dome ($R_c$) and it is given by $v_g={-dR_c(t)/dt}$, with  $R_c=\left(\left({V_{active}-N\rho_{cubic}^{-1}}\right)/{\pi l_c}\right)^{\frac{1}{2}}$ where $V_{active}$ is the volume of the active region, $N$ is the number of crystalline atoms in  the crystallite, $\rho_{cubic}$ is the density of cubic phase and $l_c$ is the length of the  semi-cylindrical dome. On the other hand, $v_{\rm g}$ of the  crystallites nucleated inside the dome can be estimated by assuming a spherical shape from  $v_g={dR(t)/dt}$ where the radius $R$ is given by $R=\left({3N}/{4\pi\rho_{cubic}}\right)^{\frac{1}{3}}$. The evolution of $R_c(t)$ as a function of time in simulation A is reported in  Figure S1 in the Supporting Information, along with the radius $R(t)$ of the six largest crystallites with a number of atoms higher than 5000 (see Figure \ref{CrystallineAtoms}b).  For the crystallite growing from the rim, the resulting $v_{\rm g}$  of 7.1 m/s is very close to the value obtained for the growth at 600 K of a crystalline slab in contact with the liquid (heterogeneous crystallization) in our previous work (see Figure 6a in Ref. \cite{omar2024}) while for six largest crystallites nucleated  inside the dome the average $v_{\rm g}$ of 2.85 m/s (see  Figure S1 in the Supporting Information) is slightly lower than the values reported for homogeneous crystallization in Ref. \cite{omar2024}.

We remark that the competition between nucleation and growth from the rim is dependent on temperature because the nucleation rate decreases very rapidly above 600 K \cite{omar2024}. The latent heat released mostly in the outer crystallite growing from the rim is expected to raise the temperature inside the dome and then to reduce the nucleation rate. In  simulation A described above, we mimic the extreme condition in which the latent heat is fully removed by a fast heat diffusion due to electrons in the supercooled liquid phase. Indeed, in simulation B in which the latent heat is removed only by a thermostat in the upper part of the model, we observed an increase of temperature inside the dome during the crystallization process and then the nucleation of a lower number  of crystallites. The evolution in time of the fraction of the recrystallized atoms and distribution of the size of the crystallites are shown in Figure \ref{CrystallineAtoms}. We observed the nucleation inside the dome of 6 and about 55 crystallites in simulation B and A, respectively.  As a consequence, the size of the nuclei is larger in simulation B than in simulation A. The equivalent of Figure \ref{Snapshots} for  simulation B and a 2D map of the temperature for different simulation times are  given in Figures S2-S3 in the Supporting Information.

\begin{figure*}[]
 \centering
\includegraphics[width=0.8\textwidth, keepaspectratio]{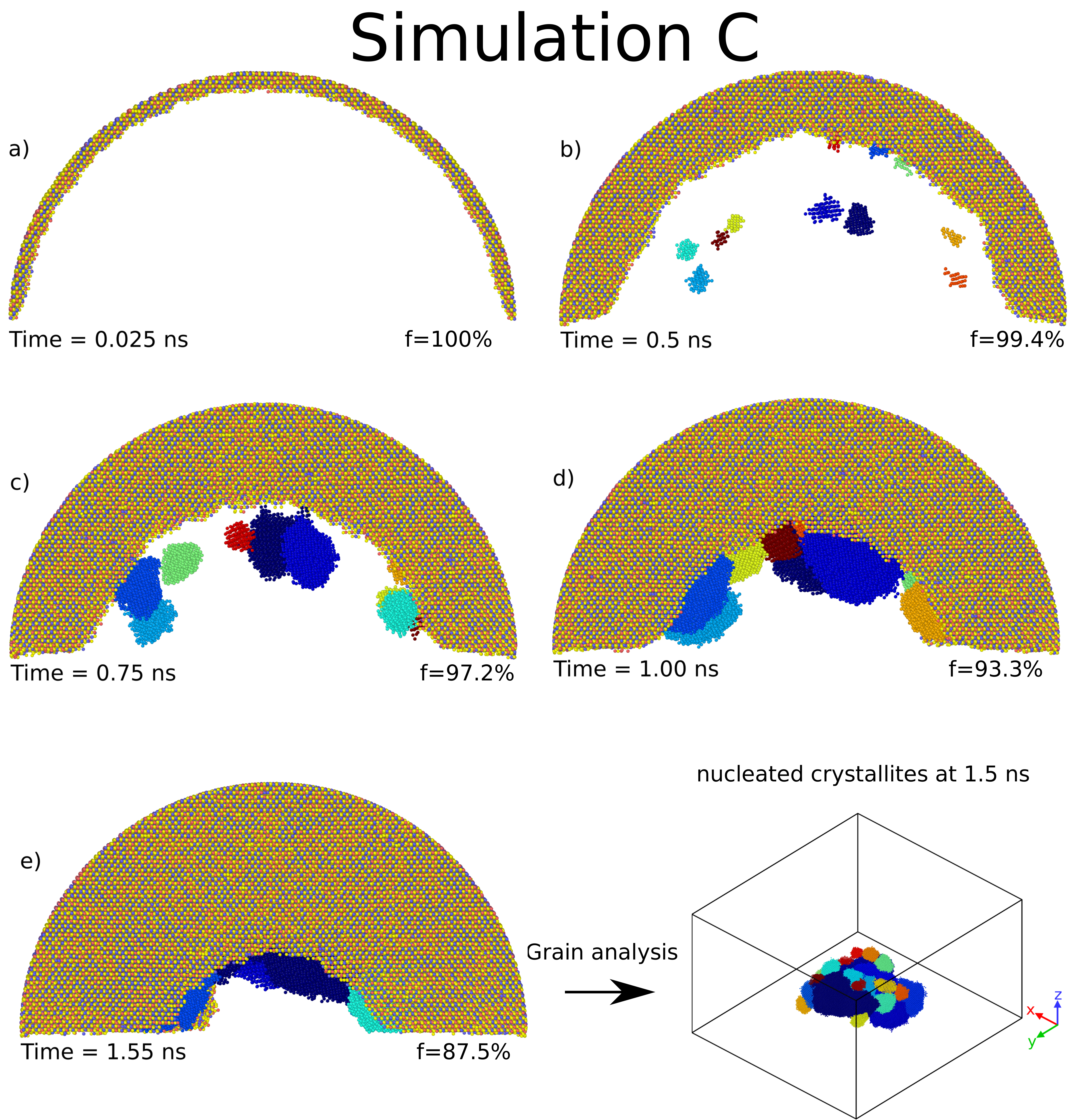}
 \caption{a-e) Snapshots of the recrystallization of the amorphous region at different times (in ns) with a thermostat in contact with the upper electrode and the hot spot inside the dome (simulation C, see Figure \ref{WallPCM}d). Only recrystallized atoms that were originally present in the amorphous semi-cylindrical dome in Figure \ref{WallPCM}b-c are shown. Atoms belonging to the largest crystallite growing form the outer amorphous-crystal interface are shown with the same colors of Figure \ref{WallPCM}c. The largest crystallite contains 554,752 atoms. The smaller crystallites nucleated inside the dome are shown each one in a different color. The fraction of crystallized atoms (number f) belonging to the outer crystallite growing from the rim is given in each panel.  A side view of the smaller crystallites (grains) embedded in the largest one is shown on bottom right. The crystallites nucleated far from the confining walls (see text) leading to the formation of a polycrystalline region at the center of the model.}
 \label{SnapshotsC}
\end{figure*}

\begin{figure*}
 \centering
 \includegraphics[width=0.7\textwidth, keepaspectratio]{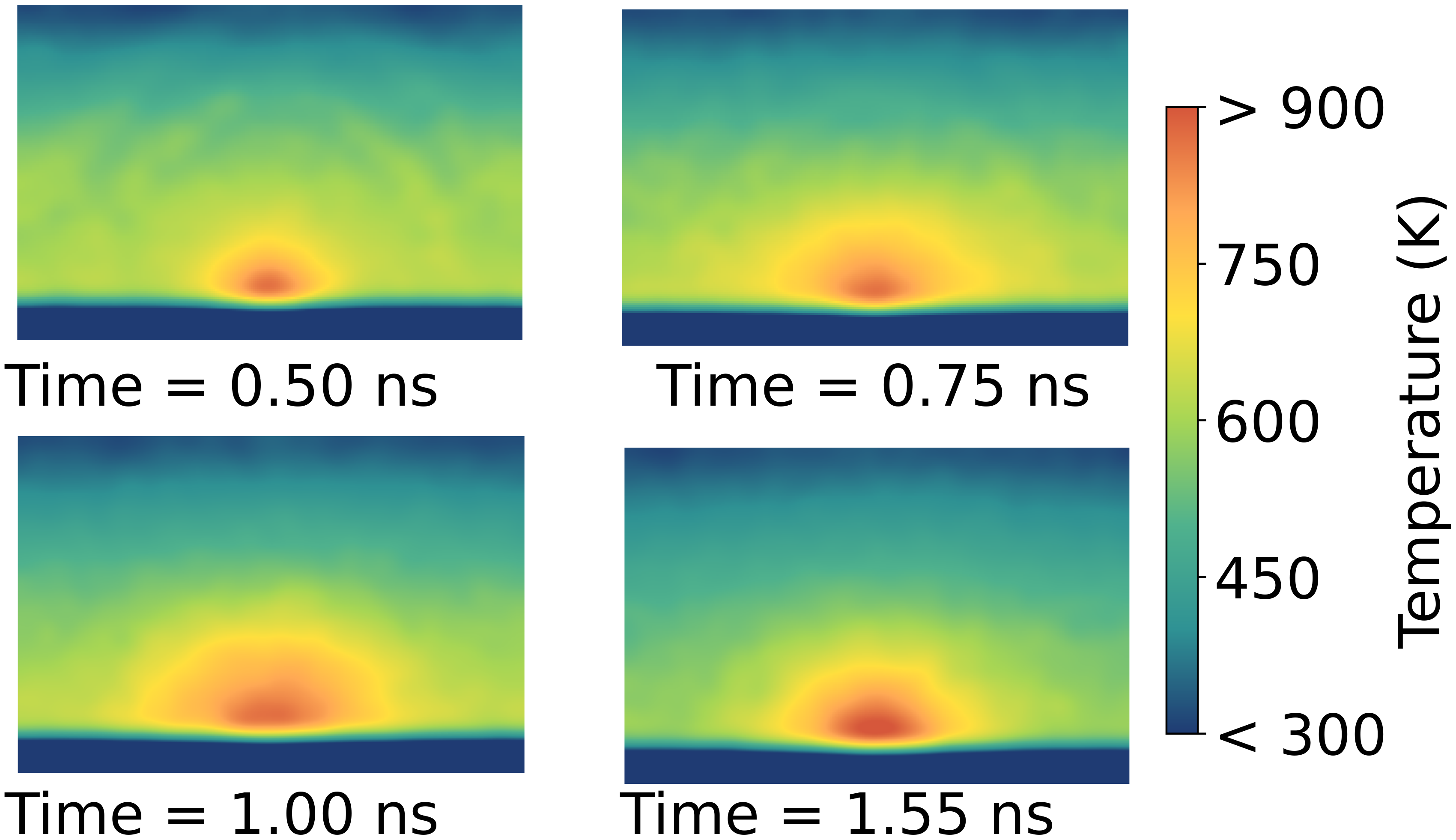}
 \caption{ A 2D map of the temperature at different times in simulation C. The $xz$ plane (see Figure \ref{WallPCM}b) was  divided into a 240$\times$165 grid (39600 pixels). The atomic temperature was then averaged over each pixel area (2$\times$2 {\AA$^2$ }) and  along the $y$ direction  in a slice 48 nm thick at the center of the model, far from the confining walls along $y$.  A Guassian filtering was then used to smear out the fluctuations.}
 \label{TmapC}
\end{figure*}

Finally, we discuss the more realistic simulation C with the hot spot inside the dome and a second thermostat in the upper part of the model. For simulation C, we report in Figure \ref{SnapshotsC} the snapshots equivalent to those in Figure \ref{Snapshots} for simulation A. The number of crystallites nucleated inside the dome is about 20, which is lower than in simulation A and larger than in simulation B. The evolution in time of the fraction of the recrystallized atoms and distribution of the size of the crystallites are shown in Figure \ref{CrystallineAtoms}.
The 2D map of the temperature for different  times in simulation C is shown in Figure \ref{TmapC}; the temperature is  mediated in the $y$ direction (see Figure \ref{WallPCM}b) in a slice 48 nm thick at the center of the model. The temperature close the confining walls is higher than at the center of the model because  the walls have an infinite thermal boundary resistance (the atoms are frozen). This effect is more evident in simulation C, because of the presence of the hot spot inside the dome.

The recrystallized system provides a very large model of the disordered   cubic phase of GST grown at the realistic time scale of the operation of the device. This model allows us to investigate the distribution of point defects, namely antisite defects (cation-anion exchange), vacancies, and vacancy clusters on both sublattices. These defects are of relevance for the electrical conductivity of the cubic crystalline phase. It was shown that clusters of vacancies on the cationic sublattice give rise to localized states at the Fermi level \cite{Mazzarello2012},  while antisite defects are responsible for the formation of  empty states close to the edge of the conduction band \cite{GST124,kostantinou}. We remark that in an ideal crystalline cubic GST with the rocksalt geometry, the face-centered-cubic (fcc) anionic sublattice if fully occupied by Te, while the fcc cationic sublattice is randomly occupied by 40 $\%$ of Ge and Sb and by 20 $\%$ of stoichiometric vacancies, i.e. ($\Box$Ge$_2$Sb$_2$)$_c$(Te$_5$)$_a$, where the subscript $a$/$c$ stand for anionic/cationic sites.  We performed the  analysis of the defects in the larger crystallite grown from the rim in  simulation A  that contains 541,233 atoms.

The largest crystallite is slightly Te-poor with a composition  Ge$_{2.021}$Sb$_{2.03}$Te$_5$ which amounts to N$_{\rm Ge}$=120,617, N$_{\rm Sb}$=121,327 and N$_{\rm Te}$=299,288 total number of Ge, Sb and Te atoms. The missing Te atoms enrich the amorphous regions between the crystalline grains. This composition arises, however, also from the presence of vacancies on both the cationic and anionic sublattice and of antisites Te$_c$, Sb$_a$ and Ge$_a$. 
The composition of the other six largest crystallites nucleated inside the dome is given in Table S1 in the Supporting Information.

To assess the reliability of the NN potential in describing antisite defects, we compared the formation energy of different defects obtained from NN and DFT calculations within the framework used to generate the NN potential itself \cite{omar2024}. We considered the 270-atom supercell of the ideal cubic GST (random occupation of the cation sublattice with Ge/Sb and 20 $\%$ of vacancies) generated in Ref. \cite{caravati2009first} by a special quasi random structure method \cite{sqs}. We then switched a Te atom with a Ge atom (Te$_c$-Ge$_a$ pair) or with an Sb atom (Te$_c$-Sb$_a$ pair) or we moved a Te atom in a cationic vacancy (Te$_{\Box}$). We considered three different switches for each type of defect.  The DFT formation energy is 1.60 $\pm$ 0.50 eV for Te$_{\Box}$, 1.35 $\pm$ 0.10 eV for the Te$_c$-Ge$_a$ pair, and 1.10 $\pm$ 0.10 eV for the Te$_c$-Sb$_a$ pair. The root-mean-square-error (RMSE) between DFT and NN formation energies is 0.28 eV for Te$_{\Box}$, 0.105 eV for the Te$_c$-Ge$_a$ pair, and 0.085 eV for the Te$_c$-Sb$_a$ pair. Overall the RMSE error is lower than the spread in the formation energy due to different environments. 
\begin{center}
\captionsetup{type=Table}
\captionof{table}{Average coordination numbers  for each chemical species  in the largest crystallite of recrystallized GST. n$_{\rm c}$ and n$_{\rm a}$ indicate the average coordination number of atoms on the cationic and anionic sublattice. The coordination in the ideal stoichiometric compound is given in parenthesis.}
\label{coord}
\begin{tabular}{c c}
\hline
\hline
      -       & Coordination \\ \hline
n$_{\rm Ge}$ &5.97 (6.0) \\
n$_{\rm Sb}$ &5.96 (6.0) \\
n$_{\rm Te}$ &4.93 (4.8) \\
n$_{\rm c}$  &5.97 (6.0) \\
n$_{\rm a}$  &4.90 (4.8) \\
 \hline
\end{tabular}
\end{center}
We then identified the antisite defects in the crystallites from the analysis of the partial coordination numbers (NCs). 

For example, a Te atom on the anionic sublattice should have NC$_{\rm TeTe}$=0, while NC$_{\rm TeTe} \neq$0 for an antisite Te atom. Therefore, we label a Te atom with NC$_{\rm TeTe}$ $\ge$ 3 as an antisite atom, by assuming that  a cationic site could have at most three vacancies as first neighbors on the anionic sites. Similarly,   Ge$_a$ and Sb$_a$ atoms were identified from the condition  NC$_{GeGe}$+NC$_{GeSb}\ge$ 3 or NC$_{\rm SbGe}$+NC$_{\rm SbSb}\ge$ 3, again assuming that an anionic site could have at most three vacancies as first neighbors on the cationic sites (see below). We found only 36 Ge$_a$, but 1606 Sb$_a$ and 3457 Te$_c$. Therefore, only a fraction of Te$_c$ arises from an exchange with cations, the majority of  Te$_c$ consists of Te atoms occupying the stoichiometric cationic vacancies. The Te$_c$, Sb$_a$ and Ge$_a$ antisites are overall 1.1, 1.3, 0.03  $\%$ of the Te, Sb, and Ge  atoms.
The number of different antisites in the six largest crystallites nucleated inside the dome is given in Table S2 in the Supporting Information.

\begin{figure}[h]
    \centering
    \includegraphics[width=0.75\linewidth]{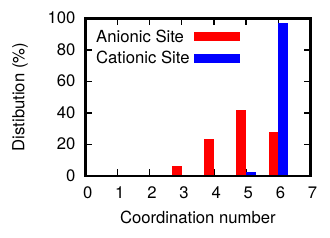} 
    \caption{Distribution of the coordination number for atoms in the anionic (red) and cationic (blue) sites of the large crystallite grown from the rim (541,231 atoms) in simulation A.}
    \label{coordist}
\end{figure}

Then, we estimated the number of vacancies  on the cationic and anionic sublattices  from the  average coordination numbers given in Table \ref{coord}. The distribution of the coordination numbers for anionic and cationic sites is shown in Figure \ref{coordist}. The number of cationic vacancies is obtained from the average coordination number of the atoms occupying anionic sublattice n$_{\rm a}$ (see Table \ref{coord}) as  $\Box_c$=(6-n$_{\rm a}$)N$_{\rm a}$/6  where the factor six comes from the presence of six nearest neighbors and N$_{\rm a}$ is the number atoms occupying  anionic sublattice (i.e. Ge$_a$, Sb$_a$ and Te$_a$). Similarly, the number of anionic vacancies is obtained by $\Box_a$=(6-n$_{\rm c}$)N$_{\rm c}$/6.  This calculation yields the following composition including antisites and vacancies on the two sublattices 
($\Box_{0.9173}$Ge$_{2.0155}$Sb$_{2.0012}$Te$_{0.0578}$)$_c$- ($\Box_{0.0277}$Ge$_{0.0006}$Sb$_{0.0268}$Te$_{4.9449}$)$_a$ to be compared with the ($\Box$Ge$_2$Sb$_2$)$_c$(Te$_5$)$_a$ formula unit for the ideal crystal with stoichiometric vacancies only.

About 10 $\%$ (0.91 vs 1.0)  of the stoichiometric vacancies on the cationic sublattice is filled by Ge, Sb and Te antisite (Te$_c$). A fraction of about 0.5 $\%$ of the anionic sublattice is also occupied by vacancies. 

The overall concentration of antisites is below 0.95 $\%$ which means that the two sublattices are very strongly chemically ordered in a crystallite grown in a few ns at 600 K. This is at odds with the recent finding of a strong chemical disorder during the growth of the metastable cubic phase of the Sb$_2$Te$_3$ which is a parent compound \cite{kolobovSbTe} of Ge$_2$Sb$_2$Te$_5$ 
that can in fact be seen as a pseudobinary alloy
along the GeTe-Sb$_2$Te$_3$ tie-line in the ternary phase diagram.

We have then made a step further by analyzing the clustering of vacancies and antisites. The clustering of vacancies in the cationic sublattice can be inferred from the distribution of the coordination numbers of atoms occupying anionic sublattice (mostly Te) as shown in Figure \ref{coordist}. Most of these atoms are 5-coordinated which means that they have a single nearest neighbor vacancy, while about than 20 $\%$ of Te atoms have two nearest neighbor vacancies (4-fold coordinated), and about 5 $\%$  have three nearest neighbor vacancies (3-fold coordinated), and only 0.4\% of Te atoms are 2-fold coordinated. Regarding the clustering of antisites, we consider two antisites belonging to the same cluster if their distance is shorter than the nearest neighbor distance on the fcc sublattice (6.1 \AA). 

\begin{figure}[]
    \centering
    \includegraphics[width=0.75\linewidth]{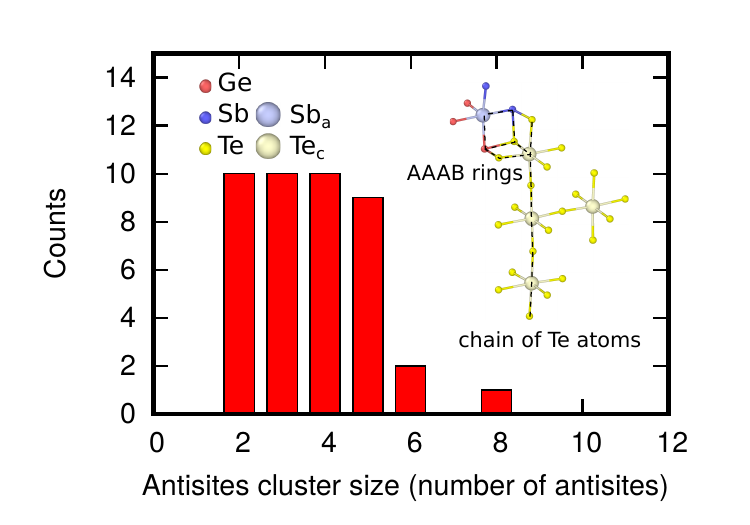} 
    \caption{Number of clusters of antisites with different sizes.  A snapshot of a 5-membered cluster is given in the inset which is made of a pair of AAAB/BBBA rings and a long chian of Te atoms.}
    \label{antisiteDist}
\end{figure}

The distribution of the size of the clusters of antisites is shown in Figure \ref{antisiteDist}. Overall 71 $\%$ of the antisites are isolated, while 1455 out of 5099 antisites are arranged in clusters of different sizes.

\begin{figure}[]
    \centering
    \includegraphics[width=0.7\linewidth,keepaspectratio]{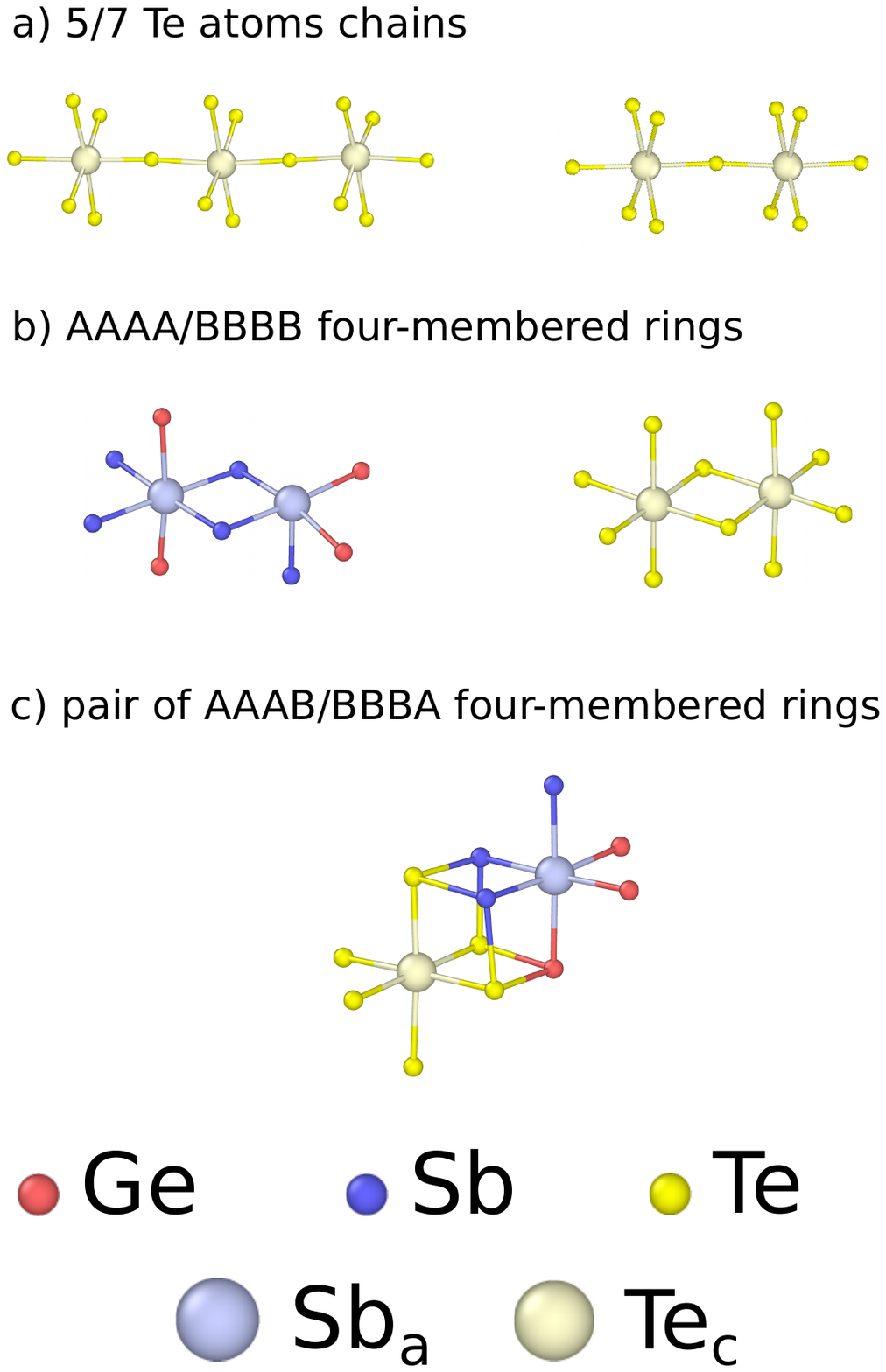}
    \caption{Snapshots of the different types of clusters of antisites.}
    \label{SnapshotAntisites}
\end{figure}

About 3\% of these clusters has a  swapped pair of adjacent atoms (Ge$_a$/Sb$_a$-Te$_c$ first neighbors), while most of the clusters with two/three antisite atoms are either linear chains of  Te atoms or defective four-membered rings, such as isolated AAAA or BBBB rings where A=Ge/Sb and B=Te, or a pair of  AAAB/BBBA rings forming a cube. Snapshots of the different types of antisite clusters are shown in Figure \ref{SnapshotAntisites}.
Overall, we observed a rather large fraction of antisite defects (0.95 $\%$ of the total number of atoms) which mostly form chains of Te-Te homopolar bonds. According to the DFT calculations in Ref. \cite{GST124,kostantinou}, these defects give rise to localized states close to the conduction band edge. Moreover, a sizable fraction of Te atoms (5 $\%$) are 3-fold coordinated due to the clustering of three vacancies on the cationic sublattice. These latter configurations are responsible for localized states at the edge of the valence band as shown by DFT calculations in Ref. \cite{Mazzarello2012,GST124}.

\section{Conclusions}
In summary, we have shown that the NN potential developed with the DeepMD code allows simulating the crystallization process of the prototypical phase change compound Ge$_2$Sb$_2$Te$_5$ at the length and time scales of the real memory devices. This first multimillion atom simulation of a model mimicking the real memory device in the Wall architecture, unveils the subtle competition between homogeneous crystal nucleation and crystal growth from the crystal/amorphous rim and its dependence on the temperature profile generated by the programming protocol. The simulations provided a vivid atomistic picture of unprecedented realism of the operation of the memory device and they pave the way to the simulation of more complex transformation at the device scale such as the crystallization and phase separation in Ge-rich GeSbTe alloys exploited in embedded memories \cite{Cappelletti_2020,Redaelli2022} and whose kinetics details  are still largely unknown. 
Moreover, the recrystallized system provided a very large model of cubic GST grown at the real time scale of the memory operation which allowed us to gain a statistically sound distribution of the defects
(antisites and vacancies)  that are known to give rise to the localized states in the band gap which control the electrical resistivity of the set state.

\section{Computational Details}
MD simulations have been performed by using the NN potential of Ref. \cite{omar2024},  generated  with the DeePMD package \cite{wang2018deepmd,PhysRevLett.120.143001,lu202186} by fitting a database DFT energies and forces of around 180.000 configurations of small supercells (57-108 atoms). The Perdew-Burke-Ernzerhof (PBE) \cite{PBE} exchange and correlation functional and norm conserving pseudopotentials were used \cite{GTH1,GTH2}. The potential was validated in Ref. \cite{omar2024} on the structural and dynamical properties of the liquid, amorphous and crystalline phases and  it was then exploited to study the crystallization kinetics in the bulk with supercells containing up to 12900 atoms. The LAMMPS code \cite{LAMMPS} has been used as MD driver with a time step of 2 fs and a Nos\'e-Hoover thermostat \cite{noseart,hoover}.

\medskip

\textbf{Data availability} \par  Trajectories of the crystallization process for the three simulations A/B/C are available upon request.

\medskip

\textbf{Acknowledgements} \par 
The project has received funding from European
Union NextGenerationEU through the Italian Ministry of University and Research under PNRR M4C2I1.4 ICSC 
Centro Nazionale di Ricerca in High Performance Computing,
Big Data and Quantum Computing (Grant No. CN00000013).

\medskip
\textbf{Code availability} \par
LAMMPS, and DeePMD are free and open source codes available at https:// lammps.sandia.gov  and http://www.deepmd.org, respectively. The NN potential and the DFT database used for its generation, as reported in Ref. \cite{omar2024}, are available in the Materials Cloud repository at https://doi.org/10.24435/materialscloud:a8-45.
\medskip

\textbf{Competing interests} \par
The authors declare no competing interests.
\medskip

\textbf{Authors contributions} \par
The authors equally contributed to the work.

\medskip

\bibliographystyle{MSP}
\bibliography{ref.bib}

\onecolumngrid
\newpage
\noindent{\large\textbf{Million-atom simulation of the set process in phase change memories at the real device scale - Supporting information}}
\setcounter{figure}{0}    
\setcounter{table}{0}   

\begin{figure}[h]
 \renewcommand\figurename{Figure~S$\!\!$}
    \centering
    \includegraphics[width=0.3\linewidth,keepaspectratio]{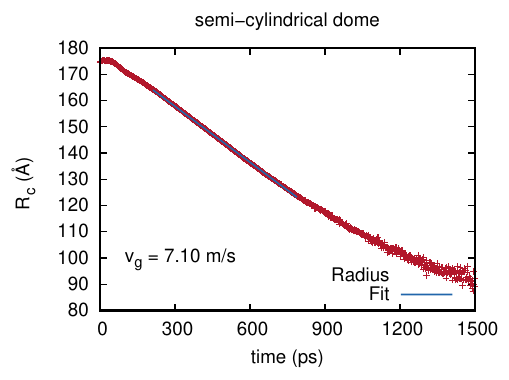}    
    \includegraphics[width=0.3\linewidth,keepaspectratio]{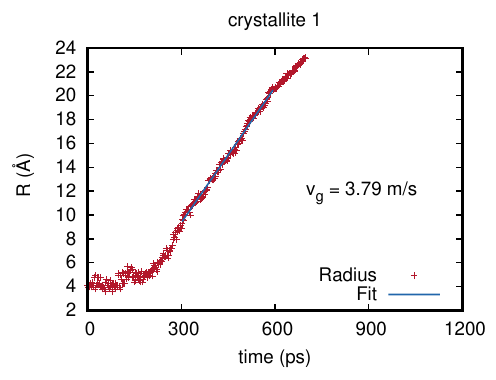}    
    \includegraphics[width=0.3\linewidth,keepaspectratio]{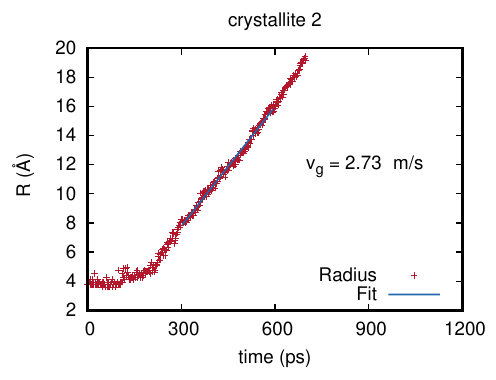}    
    \includegraphics[width=0.3\linewidth,keepaspectratio]{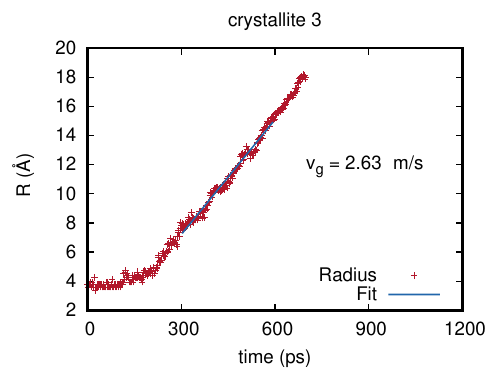}
    \includegraphics[width=0.3\linewidth,keepaspectratio]{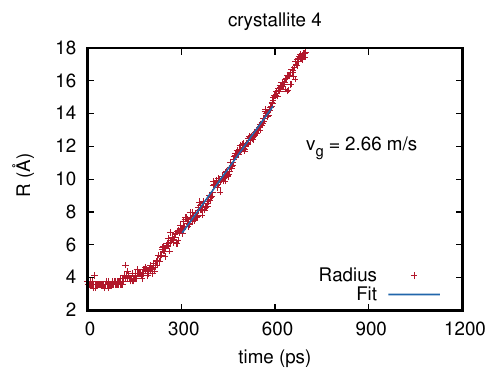}    
    \includegraphics[width=0.3\linewidth,keepaspectratio]{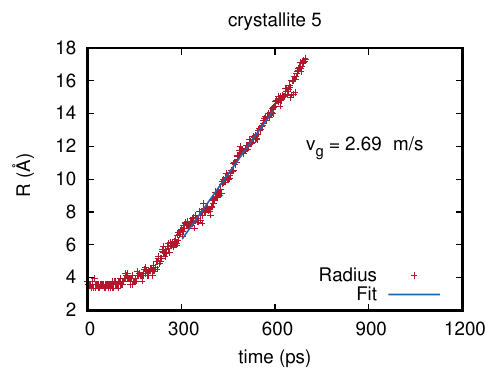}    
    \includegraphics[width=0.3\linewidth,keepaspectratio]{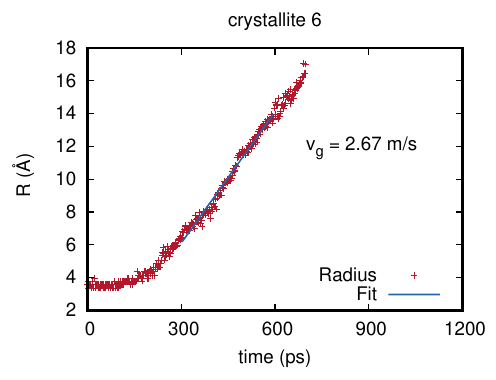}
    
    \caption{Evolution in time of the radius $R_c$ of the semi-cylindrical recrystallized dome (top left panel) and of the radius $R$ of the six largest crystallites nucleated inside the dome. The crystal growth velocity resulting from the linear fit is given as an inset.}
    \label{fig:vg}
\end{figure}

\begin{figure}[]
  \renewcommand\figurename{Figure~S$\!\!$}
 \centering
\includegraphics[width=0.8\textwidth, keepaspectratio]{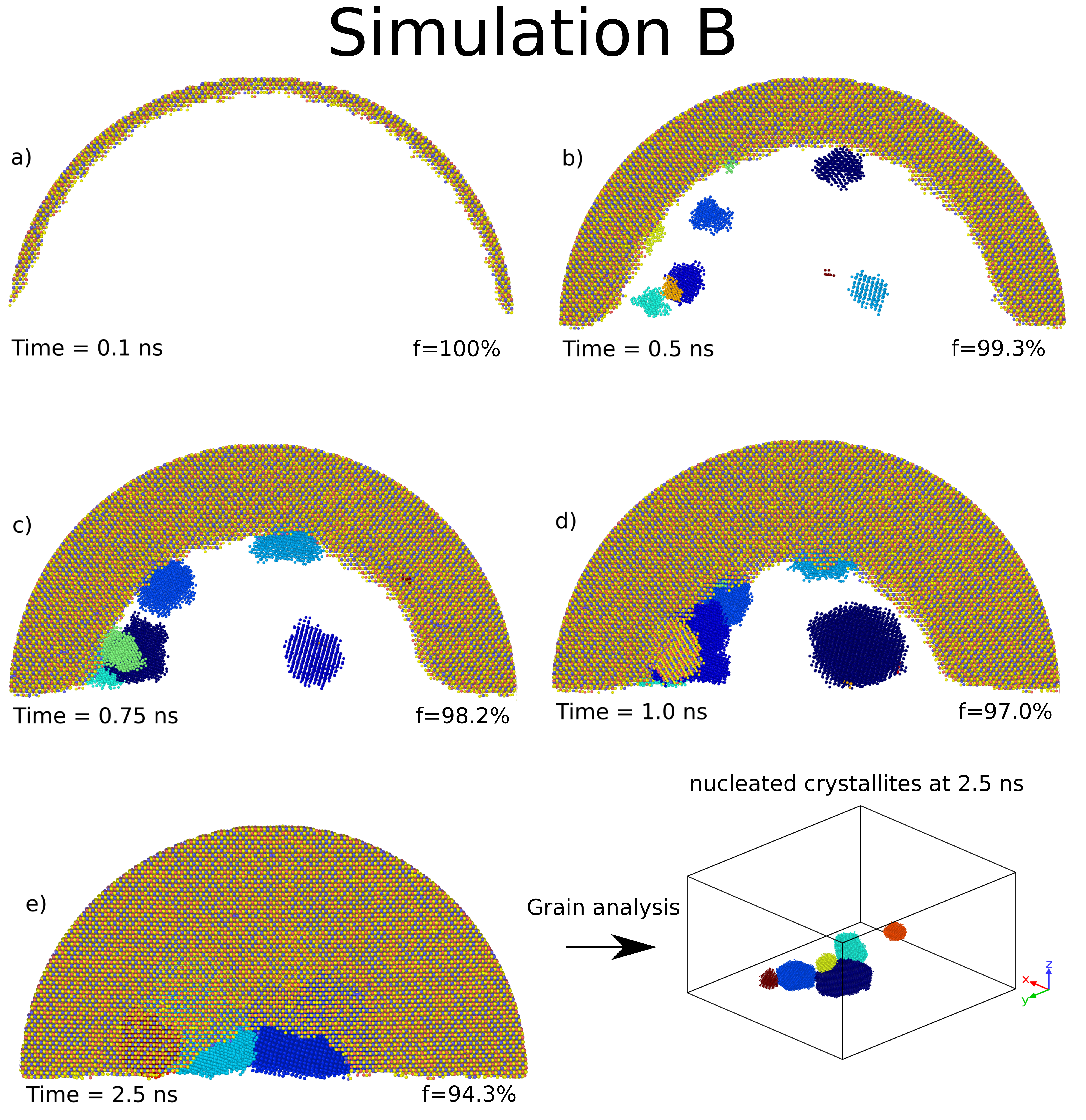}
 \caption{a-e) Snapshots of the recrystallization of the amorphous region at different times with a thermostat only in the upper part of the model mimicking the electrode (simulation B, see article). Only recrystallized atoms that were originally present in the amorphous semi-cylindrical dome (see Fig.1b-c in the article) are shown. Atoms belonging to the largest crystallite growing form the outer amorphous-crystal interface are shown with the same colors of Fig. 1c in the article. The largest crystallite containing 645158 atoms. The smaller crystallites nucleated inside the dome are shown each one in a different color. The fraction of crystallized atoms (number f) belonging to the outer crystallite growing from the rim is given in each panel.  A side view of the smaller crystallites (grains) embedded in the largest one is shown on bottom right.}
 \label{SnapshotsB}
\end{figure}

\begin{figure}[]

  \renewcommand\figurename{Figure~S$\!\!$}
 \centering
\includegraphics[width=0.75\textwidth, keepaspectratio]{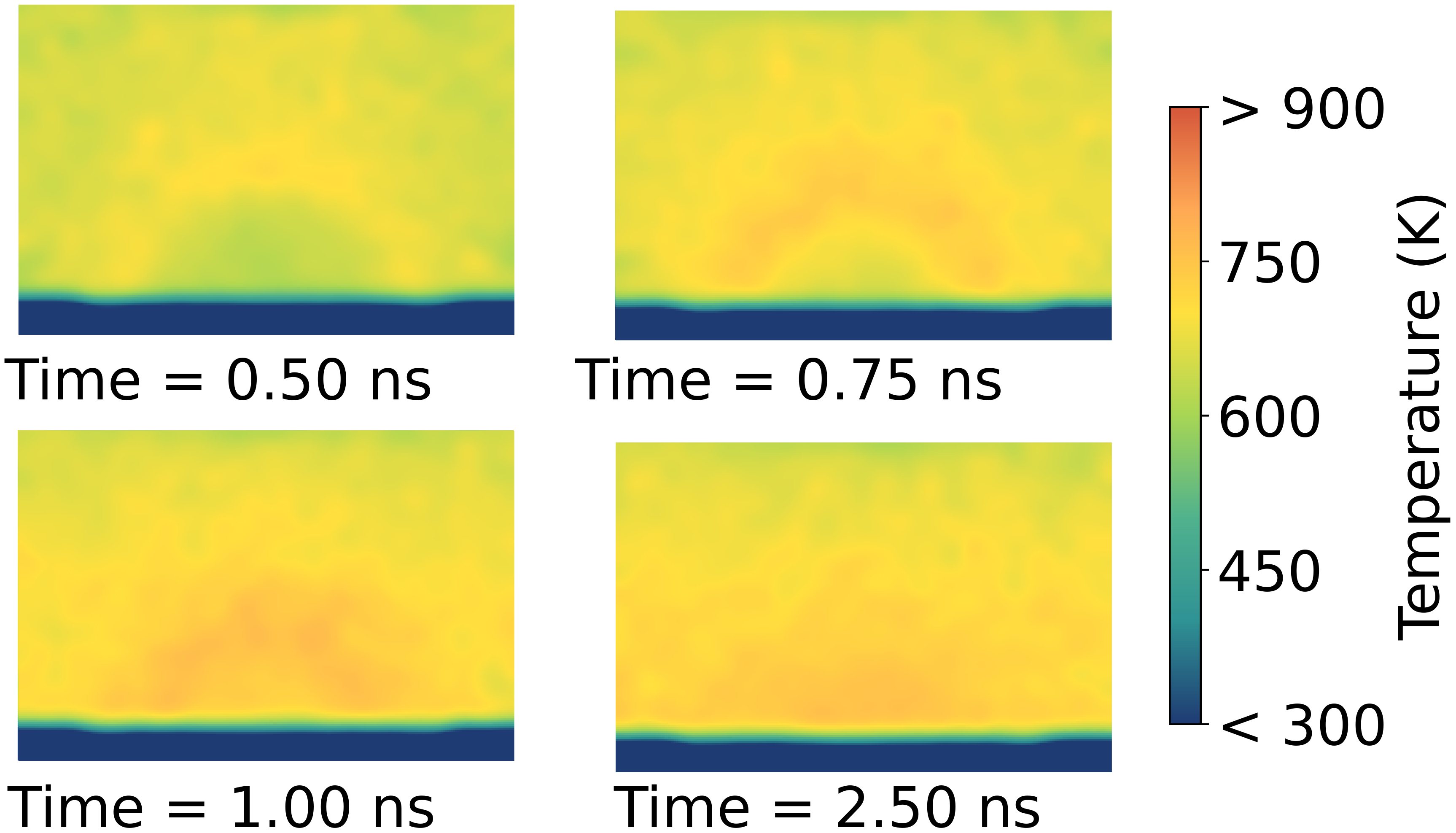}
 \caption{
 A 2D map of the temperature at different times in simulation B.
 The $xz$ plane (see Fig. 1b in the article) was first divided into a 240$\times$165  grid (39600  pixels). The atomic temperature was then averaged over each pixel area (4$\times$2 {\AA$^2$ }) and  along the $y$ direction  in a slice 48 nm thick at the center of the model, far from the confining walls along $y$.  A Guassian filtering was then used to smear out the fluctuations.}
 \label{TmapB}
\end{figure}

\newpage
\begin{table}[]
  \renewcommand\tablename{Table~S$\!\!$}
\centering
\caption{Composition of the seven largest crystallites in simulation A (see article).}
\begin{tabular}{c c c }
\hline
\hline
Crystallite index & Average Composition & Size (Atoms)\\
\hline
2992881&Ge$_{2.01507}$Sb$_{2.02693}$Te$_5$&541232\\
2&Ge$_{2.06546}$Sb$_{2.05643}$Te$_5$&12123\\
3&Ge$_{2.09951}$Sb$_{2.08524}$Te$_5$&9655\\
4&Ge$_{2.05752}$Sb$_{2.08344}$Te$_5$&7406\\
5&Ge$_{2.13184}$Sb$_{2.03787}$Te$_5$&6538\\
6&Ge$_{2.13599}$Sb$_{2.08066}$Te$_5$&6330\\
7&Ge$_{2.0496 }$Sb$_{2.0983 }$Te$_5$&6012\\
Average &Ge$_{2.02089}$Sb$_{2.03058}$Te$_5$&589296\\

\hline

\end{tabular}
\end{table}

\begin{table}[]
  \renewcommand\tablename{Table~S$\!\!$}
\centering
\caption{Number of atoms of the different species, and number of the different antisites in the seven largest crystallites from simulation A (see article).}
\begin{tabular}{c c c c c c c}
\hline
\hline

Crystallite index   &\# Ge  &\# Sb   &\# Te      &\# Ge$_{a}$   &\# Sb$_{a}$     &\# Te$_{c}$\\  
\hline
1 &     120581  &119721 &295831  &36    &1606   &3457\\
2 &     2745    &2718   &6625    &0     &15     &20\\  
3 &     2206    &2181   &5242    &1     &11     &14\\  
4 &     1667    &1680   &4041    &0     &8      &10\\  
5 &     1520    &1448   &3561    &0     &5      &4\\   
6 &     1467    &1418   &3432    &0     &11     &2\\   
7 &     1347    &1376   &3279    &0     &3      &7\\   
Total:& 131533  &130542 &322011  &37    &1659   &3514\\

\hline

\end{tabular}
\end{table}

\end{document}